\documentclass[12pt]{article}
\usepackage{amsmath}
\usepackage{graphics,graphicx,epsfig}
\usepackage{amssymb}
\usepackage{epstopdf}
\usepackage{sidecap}
\usepackage{appendix}

\def\tvphi{\tilde\varphi}

\begin{document}

\begin{titlepage}
\vfill
\begin{flushright}
ACFI-T13-03
\end{flushright}

\vfill
\begin{center}
\baselineskip=16pt
{\Large\bf Magnetic Fields in an Expanding Universe}
\vskip 0.15in
\vskip 10.mm
{\large\bf 
David Kastor\footnote{kastor@physics.umass.edu} and Jennie Traschen\footnote{traschen@physics.umass.edu}} 

\vskip 0.5cm
{{Amherst Center for Fundamental Interactions\\
Department of Physics, University of Massachusetts, Amherst, MA 01003
     }}
\vspace{6pt}
\end{center}
\vskip 0.2in
\par
\begin{center}
{\bf Abstract}
 \end{center}
\begin{quote}
We find a solution to $4D$ Einstein-Maxwell theory coupled to a massless dilaton field describing a Melvin magnetic field in an expanding universe with `stiff matter' equation of state parameter $w=+1$.  As the universe expands, magnetic flux becomes more concentrated around the symmetry axis for dilaton coupling $a<1/\sqrt{3}$ and more dispersed for $a>1/\sqrt{3}$.  An electric field circulates around the symmetry axis in the direction determined by Lenz's law.  For $a=0$ the magnetic flux through a disk of fixed comoving radius is proportional to the proper area of the disk.  This result disagrees with the usual expectation based on a test magnetic field that this flux should be constant, and we show why this difference arises.  We also find a Melvin solution in an accelerating universe with $w=-7/9$ for a dilaton field with a certain exponential potential.  Our main tools are simple manipulations in $5D$ Kaluza-Klein theory and related solution generating techniques.  We also discuss a number of directions for possible extensions of this work.
  \vfill
\vskip 2.mm
\end{quote}
\hfill
\end{titlepage}

\section{Introduction}

Magnetic fields are observed  by astronomers on scales ranging from individual stars, through galactic scales, clusters of galaxies and beyond (see {\it e.g.} the recent reviews \cite{Durrer:2013pga,Kunze:2013kza}).  They are important in a wide variety of astrophysical and cosmological phenomena.  In general relativity, the Melvin solution \cite{Melvin:1963qx} describes  the curved spacetime geometry of a static magnetic field, aligned in a fixed direction and spread throughout space.   Melvin magnetic fields have been widely applied in the context of black hole physics.
Solutions for black holes immersed in Melvin backgrounds may be obtained from non-magnetic solutions using  generating methods \cite{Ernst3,ErnstWild}\footnote{See also reference \cite{Astorino:2013xc} for a generalization to black hole-type solutions including scalar fields in Melvin magnetic backgrounds.}.   
For example, following earlier work in \cite{Galtsov:1978ag,Aliev:1980et,Dokuchaev:Jun 1987jua,Aliev:1988wy,Aliev:1989sw,Aliev:1989wz}
 detailed studies of rotating black holes, uncharged and charged, in Melvin backgrounds have been carried out recently  in \cite{Gibbons:2013yq,Gibbons:2013dna,Cvetic:2013roa}.   In a more exotic setting, the Ernst spacetime \cite{Ernst4} describes a pair of oppositely charged magnetic black holes accelerated apart in a  Melvin magnetic field.  The semiclassical decay rate of the magnetic field due to black hole pair production is given to leading order by the value of the gravitational action for the Ernst spacetime \cite{Garfinkle:1990eq}.  
In higher dimensions, Melvin magnetic fields have also recently been used to stabilize numerical solutions for dipole black rings  \cite{Kleihaus:2013zpa}.
 
Given the many uses of  Melvin magnetic fields in black hole physics, it would be interesting to have exact solutions for gravitating magnetic fields in cosmological settings as well.  We will see in this paper that it is, in fact, straightforward to add Melvin magnetic fields to certain scalar field cosmologies.  
We will initially consider solutions to $4D$ dilaton gravity 
\begin{equation}\label{action}
S=\int d^4x\sqrt{-g}\left(R-2(\nabla\phi)^2-e^{-2a\phi}F^2 \right),
\end{equation}
and later add a dilaton potential term. 
Certain values of the dilaton coupling $a$ have particular significance.
Taking $a=0$ gives ordinary Einstein-Maxwell theory coupled to a massless scalar field, while $a=1$ arises in string theory.  We will be particularly interested in the cases $a=\sqrt{3}$ corresponding to the Kaluza-Klein reduction of $5D$ vacuum Einstein gravity and $a=1/\sqrt{3}$ which comes from the reduction of $5D$ Einstein-Maxwell theory.  

Homogeneous and isotropic scalar field cosmologies with no magnetic fields, and also static Melvin type solutions to dilaton gravity, are individually well known.
A flat FRW spacetime with  scale factor $\tau^{1/3}$ (where $\tau$ is the FRW proper time coordinate)  and time dependent dilaton is a solution for all values of the dilaton coupling.  Dilatonic generalizations of the Melvin spacetime were found in \cite{Gibbons:1987ps}.  In this work, we will find cosmological Melvin solutions for all values of the dilaton coupling that combine the key features of these two types of solutions.  A  particularly simple construction holds for dilaton coupling $a=\sqrt{3}$.  In this case, the cosmological and dilaton Melvin spacetimes can each be obtained from different  Kaluza-Klein reductions of $5D$ flat spacetime in appropriate coordinates.  We show that these coordinate choices can be melded to give a cosmological Melvin spacetime with $a=\sqrt{3}$.  A similarly simple construction works for $a=1/\sqrt{3}$, while the case of general dilaton coupling can be handled via a solution generating method developed in \cite{Dowker:1993bt}.  

The evolution of the cosmological Melvin spacetimes differs dramatically between theories with dilaton coupling greater than, less than, or equal to the critical value $a_{c}=1/\sqrt{3}$.  For the critical coupling $a=a_c$ the magnetic flux disperses gradually along with the expansion of the universe.  For $a>a_c$ it disperses more rapidly with time, while for $a<a_c$ it actually becomes more concentrated as the universe expands.   This latter behavior includes the case  $a=0$ of particular physical interest.   For $a=0$ we find that the flux through a disk of constant comoving radius grows like the proper area of the disk.  This disagrees with the usual expectation, based on test fields, that the flux through such a disk should remain constant despite its growing area.  We resolve this is apparent contradiction by examining the test field limit of the cosmological Melvin spacetimes, which we find corresponds to a different solution to the Maxwell equations than is commonly assumed.  

We also examine the flux of stress energy in the cosmological Melvin spacetimes.  For the Maxwell field we find this to be directed inward (outward) for $a<a_c$ ($a>a_c$), in accordance with the behavior of the magnetic flux.  The stress-energy flux for the dilaton field is more complicated.  For $a<1/a_c$ it is always directed inward, while for $a>1/a_c$ it is directed inward or outward depending on whether the radius is greater than or less than a critical radius that increases with time.

The paper proceeds as follows.  In section (\ref{basic}) we review the dilaton Melvin spacetimes and their key features. In section (\ref{construct}) we use  $5D$ Kaluza-Klein manipulations that we call `Milnization' and `Melvinizing' to construct two simple examples of $4D$ cosmological Melvin solutions with $a=\sqrt{3}$ and $a=1/\sqrt{3}$.  We then apply a solution generating method to construct  cosmological verisons of the Melvin solutions (\ref{melvin}) for all values of the dilaton coupling $a$.  In section (\ref{properties}) we analyze the key physical properties of the cosmological Melvin spacetimes, focusing on the time evolution of the magnetic flux and flow of stress-energy.  In section (\ref{fast}) we  again use simple $5D$ Kaluza-Klein manipulations to construct a $4D$ Melvin solution with accelerated cosmological expansion sourced by a dilaton with an exponential potential.  Finally, in section (\ref{conclude}) we offer some directions for further work on magnetic fields in cosmological backgrounds.

\section{Dilaton Melvin spacetimes}\label{basic}

The generalization of the Melvin spacetime to dilaton gravity was presented in \cite{Gibbons:1987ps} (see also \cite{Dowker:1993bt}) and has the metric, gauge and scalar fields 
\begin{eqnarray}\label{melvin}
ds_4^2 &=& f^{{2\over 1+a^2}}(-dt^2 +dz^2+dr^2) + f^{-{2\over 1+a^2}}r^2d\varphi^2,\qquad A_\varphi = {Br^2\over 2f}\\
e^{-2a\phi}&=&f^{{2a^2\over 1+a^2}},\qquad f=1+{(1+a^2)B^2r^2\over 4}\nonumber
\end{eqnarray}
The original Melvin solution \cite{Melvin:1963qx} is obtained by setting $a=0$.  For $a\neq 0$ the magnetic flux in the $z$-direction acts as a source for the dilaton.  The flux through a circle of radius $r$ surrounding the $z$-axis  is given by
\begin{equation}\label{flux}
\Phi(r) = \int A_\varphi d\varphi  = {\pi Br^2\over 1+{(1+a^2)\over 4}B^2r^2}
\end{equation}
For small radius, we see that $\Phi\simeq B\pi r^2$ showing that $B$ is the strength of the magnetic field on the axis.  We also see from (\ref{flux}) that the total magnetic flux through a slice of constant $z$  is finite and given by
\begin{equation}\label{totalflux}
 \Phi = {4\pi\over (1+a^2)B}.   
\end{equation}

Another important feature of the Melvin spacetimes is how the circumference of circles depends on the radial coordinate $r$, which is given by
\begin{equation}
C(r) = {2\pi r\over (1+{(1+a^2)B^2r^2\over 4})^{2\over 1+a^2}}.
\end{equation}
For $r\ll 1/B$ the circumference of a circle grows approximately as $C\sim 2\pi r$.  
However for larger radii this growth slows and for $a<1$ the circumferences of large circles shrink back to zero.  This asymptotic closing up of circles, which in particular holds for the original Melvin spacetime with $a=0$, is similar to the behavior of onion layers.
\begin{SCfigure}
%\begin{center}
\includegraphics[width=0.65\textwidth]{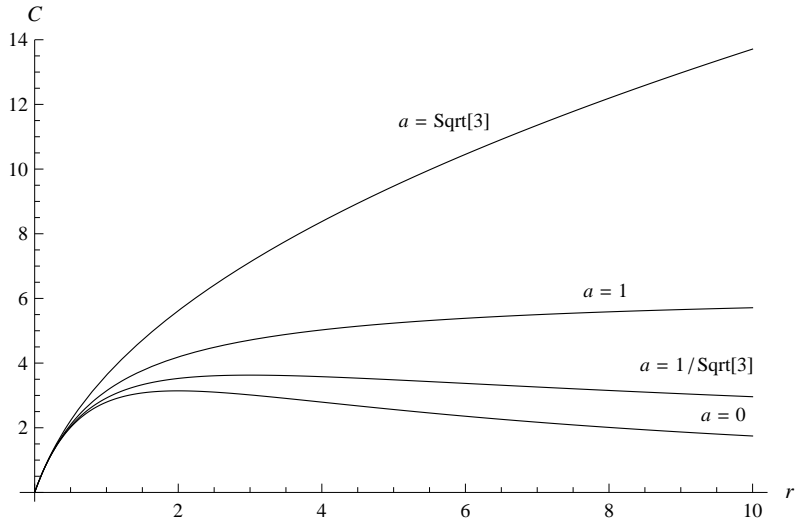}
%\end{center}
\caption{{\normalsize Circumference $C$ of circles as a function of radius $r$ for $B=1$ and for dilaton couplings $a=0,1/\sqrt{3},1$ and $\sqrt{3}$.  For $a<1$ circles shrink back down to zero circumference at large radius, while for $a=1$ they asymptote to a constant circumference and for $a>1$ they grow but more slowly than $C=2\pi r$.}} 
\label{circlefig}
\end{SCfigure}
Figure (\ref{circlefig}) shows the circumference of circles as a function of coordinate radius $r$ for a number of values of the dilaton coupling constant.
Note, however, that if the circles are extended into cylinders in the $z$-direction then factors of $f$ from the metric (\ref{melvin}) will cancel and the areas of these cylinders will be proportional to $r$ as in flat space-time.

\section{Generating cosmological Melvin spacetimes}\label{construct}

To find a cosmological version of the dilaton Melvin solutions (\ref{melvin}) we focus first on the Kaluza-Klein case.  Dilaton gravity  in $4D$ with $a=\sqrt{3}$ can be obtained from vacuum gravity in $5D$, where 
the $5D$ metric is related to the $4D$ metric $g_{ab}$, gauge and scalar 
fields appearing in the action (\ref{action}) via 
\begin{equation}\label{reduce}
ds_5^2 = e^{-4\phi/\sqrt{3}}(dw +2A_b dx^b)^2 +e^{+2\phi/\sqrt{3}}g_{ab}dx^adx^b
\end{equation}
where $a,b$ are $4D$ coordinate indices.  It is known that both magnetic fields and cosmological backgrounds in $4D$ can each individually be obtained from flat $5D$ spacetime by making appropriate identifications of points.  The new feature here will be combining these two effects to obtain cosmological Melvin spacetimes.
We then extend these results to all values of the dilation coupling  by making use of a solution generating method presented in  \cite{Dowker:1993bt}.  This will include for $a=0$ a cosmological generalization of the original Melvin solution \cite{Melvin:1963qx}.

\subsection{Kaluza-Klein Milne solution}

It is well known that a cosmological solution to dilaton gravity (\ref{action}) can be obtained by the Kaluza-Klein dimensional reduction of flat  $5D$ spacetime via a process we will call  `Milnization'.  Start with Minkowski spacetime 
\begin{equation}
ds_5^2 = -d\tilde t^2 +d\tilde w^2 +dz^2 +dr^2 +r^2d\tilde\varphi^2
\end{equation}
and transforms to Milne coordinates $(t,w)$ via
\begin{equation}\label{milnetransform}
\tilde t=t\cosh w,\qquad \tilde w=t\sinh w
\end{equation}
so that the metric now has the form
\begin{equation}\label{milneform}
ds_5^2 = -dt^2 + t^2dw^2 +dr^2+dz^2+r^2 d\tvphi^2.
\end{equation}
Making the usual KK identification $w\equiv w+2\pi L$ and dimensionally reducing according  to (\ref{reduce}) one finds that the $4D$ metric and dilaton are given by
\begin{equation}\label{4milne}
ds_4^2 = t\left( -dt^2 +dr^2+dz^2+r^2 d\tvphi^2\right),\qquad
e^{-4\phi/\sqrt{3}}= t^2. 
\end{equation}
Although  the $5D$ spacetime (\ref{milneform}) decompactifies at late times, the solution (\ref{4milne}) is nonetheless valid for all times as a solution to the $4D$ dilaton gravity theory (\ref{action}).
Transforming to ordinary, non-conformal FRW time, the spacetime (\ref{4milne}) becomes
\begin{equation}\label{FRWmilne}
ds_4^2 = -d\tau^2 +\tau^{2/3}(dr^2+dz^2+r^2 d\tvphi^2),\qquad
e^{-4\phi/\sqrt{3}}= \tau^{4/3}
\end{equation}
Interpreting the dilaton stress-energy in terms of a perfect fluid equation of state $p=w\rho$, this corresponds to `stiff matter' with $w=+1$.  Note that although we used Kaluza-Klein theory to obtain this solution, since the gauge field vanishes, it is simply a solution to $4D$ Einstein gravity coupled to a massless scalar field, and in particular to dilaton gravity for any value of the dilaton coupling $a$.  Note also that Milnization may be applied to other $5D$ spacetimes that are boost symmetric in the $(\tilde t,\tilde w)$-plane and independent of both coordinates\footnote{Such constructions may also be viewed as time dependent orbifolds.  This is commonly done in studies of string theory in cosmological backgrounds (see {\it e.g.} the review \cite{Cornalba:2003kd}).}.

\subsection{Kaluza-Klein Melvin solution}\label{melvinizing}

As shown in  \cite{Dowker:1993bt,Dowker:1995gb} the Kaluza-Klein Melvin solution given by (\ref{melvin}) with  $a=\sqrt{3}$ can also be obtained via a dimensional reduction of flat $5D$ spacetime via a procedure we call `Melvinizing'.  One starts again with
\begin{equation}\label{flat}
ds_5^2 = -dt^2 +d w^2 +dz^2 +dr^2 +r^2d\tilde\varphi^2
\end{equation}
but now with points identified according to
\begin{equation}\label{kkmelvin}
w \equiv  w+2\pi n_1 L\qquad
\tvphi \equiv  \tvphi + 2\pi n_1 LB +2\pi n_2 ,
\end{equation}
where $n_1$ and $n_2$ are integers.  Setting the parameter $B=0$ gives the usual Kaluza-Klein vacuum, resulting in a flat $4$-dimensional metric, together with vanishing scalar and electromagnetic fields.  Taking $B\neq 0$ `twists' the Kaluza-Klein circle, so that traveling around the KK direction $w$ is accompanied by a rotation around the $z$-axis in the non-compact space.
One can then transform to a new azimuthal coordinate $\varphi=\tvphi -Bw$ which together with $w$ have the simple identifications
\begin{equation}
w \equiv  w+2\pi n_1 L,\qquad
\varphi \equiv \varphi +2\pi n_2 .
\end{equation}
In terms of the new azimuthal coordinate, the flat $5D$ metric (\ref{flat}) is now given by
\begin{equation}
ds_5^2 = (1+B^2r^2)(dw+{Br^2\over 1+B^2r^2}d\varphi)^2 -dt^2 +dr^2+dz^2+{r^2\over 1+B^2\varphi^2} d\tvphi^2
\end{equation}
The $4D$ fields for the Kaluza-Klein Melvin solution can now be read off from (\ref{reduce}) to be
\begin{eqnarray}\label{newmelvin}
ds_4^2 &=& f^{1/2}\left(-dt^2 +dz^2+dr^2\right) +{r^2\over f^{1/2}}d\varphi^2 ,\qquad   A_\varphi = {Br^2\over 2f}\\
e^{-4\phi/\sqrt{3}}&=& f    ,\qquad f=1+B^2r^2\nonumber
\end{eqnarray}
which reproduces the dilaton Melvin solution (\ref{melvin}) for $a=\sqrt{3}$.  Note that shifting $B\rightarrow B+1/L$ gives the same set of identifications (\ref{kkmelvin}) of points in $5D$ \cite{Dowker:1995gb}.  However, regarded as $4D$ spacetimes, the solutions (\ref{newmelvin}) are distinct for all values of $B$.  

Note also that other $5D$ spacetimes  may also be Melvinized yielding  solutions to $4D$ dilaton gravity with $a=\sqrt{3}$ that include Melvin-type magnetic backgrounds.  For example, the Kaluza-Klein Ernst metric was obtained in this way \cite{Dowker:1993bt}.

\subsection{Cosmological Kaluza-Klein Melvin}

It is now straightforward to combine the processes of Milnization and Melvinizing by adding a magnetizing twist (\ref{kkmelvin}) to flat $5D$ spacetime in Milne form (\ref{milneform}).  The change of coordinates $\varphi=\tvphi -Bw$ and dimensional reduction (\ref{reduce}) now yield a new cosmological KK Melvin solution to dilaton gravity with $a=\sqrt{3}$
\begin{eqnarray}\label{kkcosmomelvin}
ds_4^2 &=& t\left(f^{1\over 2}(-dt^2 +dz^2+dr^2 )+f^{1\over 2}r^2d\varphi^2\right),\qquad A_\varphi = {Br^2t^{-2}\over 2f}\\
e^{-2\sqrt{3}\phi}&=& t^3f^{3\over 2},\qquad f=1+B^2r^2t^{-2}.\nonumber
\end{eqnarray}
This reduces to the scalar cosmology (\ref{4milne}) for $B=0$.  We will postpone a discussion of the properties of this solution until section (\ref{properties}).  However, note that the time and space dependence of the gauge potential $A_\varphi$ yields both time varying magnetic fluxes and electric fields in accordance with Faraday's law.

\subsection{Starting from $5D$ Einstein-Maxwell theory}\label{second}

A cosmological version of the dilaton Melvin spacetime (\ref{melvin}) with coupling $a=1/\sqrt{3}$ may also be constructed using Kaluza-Klein theory.  This value of the dilaton coupling arises from the dimensional reduction of $5D$ Einstein-Maxwell theory \cite{Gibbons:1994vm}.   Starting from the $5D$ action
\begin{equation}\label{EM5}
S_5= \int d^5x \sqrt{-G} \left(R^{(5)}-G^{AC}G^{BD}F_{AB}F_{CD}\right),
\end{equation}
where $G_{AB}$ denotes the $5D$ metric, and dimensionally reducing using equation (\ref{reduce}) with the Kaluza-Klein gauge field set to zero yields the $4D$ dilaton gravity action (\ref{action}) with $a=1/\sqrt{3}$.   This means that we can lift the static, dilaton Melvin solution (\ref{melvin}) with $a=1/\sqrt{3}$ to a solution of (\ref{EM5}) in $5D$, which gives
\begin{equation}\label{new5}
ds_5^2 = f (-d\tilde t^2 + d\tilde w^2 +dz^2 +dr^2) +{r^2d\varphi^2\over f^2},\qquad A_\varphi={Br^2\over 2f},\qquad f=1+{B^2r^2\over 3}
\end{equation}
In the terminology of \cite{Costa:2000nw,Gutperle:2001mb,Costa:2001ifa} this is an $F2$-brane, with $2+1$ dimensions tangent to the fluxbrane.  The metric is Lorentz invariant in the $(\tilde t,\tilde w)$-plane and therefore a candidate for Milnization.  Applying the coordinate transformation (\ref{milnetransform}) yields the $5D$ metric
\begin{equation}
ds_5^2 = f (-d t^2 + t^2 d w^2 +dz^2 +dr^2) +{r^2d\varphi^2\over f^2}
\end{equation}
with the function $f$ and the gauge potential still as given in (\ref{new5}).
Note that the coordinate transformation does not introduce any time dependence into the function $f$ or the gauge potential.  Dimensionally reducing using (\ref{reduce}) and setting  $A^{(4)}_\varphi=A^{(5)}_\varphi$ now gives the new cosmological Melvin solution with $a=1/\sqrt{3}$
\begin{equation}\label{another}
ds^2_4 = t\left( f^{3/2}(-dt^2 +dz^2 +dr^2) +{r^2d\varphi^2\over f^{3/2}}\right),\quad e^{-2\phi/\sqrt{3}}=f^{1/2}t,
\end{equation}
with $A_\varphi$ and $f$ given in (\ref{new5}).  Because the function $f$ is time independent,  the electric field vanishes.  We will see below that this behavior for $a=1/\sqrt{3}$ is a borderline case between qualitatively different behaviors for smaller and larger values of the dilaton coupling.

\subsection{General dilaton coupling}
Reference \cite{Dowker:1993bt} provided a solution generating method to add a Melvin-type magnetic field to any axisymmetric solution of (\ref{action}).  Note that the seed solution  is not required to be static.
If one starts with a solution with vanishing gauge potential, this transformation has the form
\begin{eqnarray}
g^\prime_{ij}&=&f^{{2\over 1+a^2}} g_{ij},\qquad g^\prime_{\varphi\varphi}=f^{-{2\over 1+a^2}}g_{\varphi\varphi},\qquad A_\varphi^\prime = {2\over (1+a^2)Bf}\\
e^{-2a\phi^\prime}&=&f^{{2a^2\over 1+a^2}}e^{-2a\phi},\qquad f= 1+ {(1+a^2)\over 4}B^2 g_{\varphi\varphi}e^{2a\phi}\nonumber
\end{eqnarray}
where the indices $i,j$ run over $t,r,z$.
As a seed solution we take the scalar field cosmology (\ref{4milne}).  This was originally obtained using Kaluza-Klein theory.  However,  because the gauge field vanishes, the $4D$ metric and scalar field solve the equations of motion for any value of the dilation coupling.  One then finds the cosmological dilaton Melvin solutions
\begin{eqnarray}\label{cosmomelvin}
ds_4^2 &=& t\left(f^{{2\over 1+a^2}}(-dt^2 +dz^2+dr^2) + f^{-{2\over 1+a^2}}r^2d\varphi^2 \right),\qquad A_\varphi ={Br^2t^{1-\sqrt{3}a}\over 2f}\\
e^{-2a\phi}&=&t^{\sqrt{3}a}f^{{2a^2\over 1+a^2}},\qquad f=1+{(1+a^2)\over 4}B^2r^2t^{1-\sqrt{3}a}\nonumber
\end{eqnarray}
This correctly reproduces the  $a=\sqrt{3}$ cosmological Melvin solution in (\ref{kkcosmomelvin}) and also the $a=1/\sqrt{3}$ cosmological Melvin solution in (\ref{another}), both of which were found above using simple manipulations in  Kaluza-Klein theory.
For $a=0$ one obtains a cosmological version of the original Melvin spacetime
\begin{eqnarray}\label{zerocosmomelvin}
ds_4^2 &=&t\left[f^2(-dt^2 +dz^2+dr^2)+{r^2\over f^2}d\varphi^2\right]
,\qquad A_\varphi ={Br^2t\over 2f}\\
\phi &=& -{\sqrt{3}\over 2}\ln t\nonumber
,\qquad f=1+{1\over 4}B^2r^2 t
\end{eqnarray}
in Einstein-Maxwell theory coupled to a massless scalar field.  For $B=0$ the solutions (\ref{cosmomelvin}) reduce to the massless scalar cosmologies (\ref{4milne}), which have $\tau^{1/3}$ power law expansion in terms of the FRW proper time coordinate.  This behavior continues to hold for all $a$ in the region where the function $f\simeq 1$.

\section{Properties of cosmological Melvin solutions}\label{properties}

In this section we will explore the key physical properties of the the cosmological Melvin solutions (\ref{cosmomelvin}).  We focus first on the geometry and then discuss the behavior of the electromagnetic fields and stress-energy flux.    For $B=0$ the cosmological Melvin solutions reduce to the $w=+1$ scalar field cosmology (\ref{4milne}).  For $B\neq 0$ it is therefore reasonable to expect that they describe a magnetic field in an expanding universe.  However, this identification is not entirely straightforward.  As with the static Melvin spacetimes (\ref{melvin}) the gravitational impact of the magnetic field is not spatially localized, and the cosmological Melvin metrics do not reduce to (\ref{4milne}) at large radius.
Nonetheless, at least for $a\neq 1/\sqrt{3}$, we will see that there is a regime, either at early or late times, in which the spacetimes (\ref{cosmomelvin}) do approach the homeogeneous and isotropic cosmology (\ref{4milne}).  This will correspond to eras in which the magnetic field becomes widely dispersed over space.

There are two important functions of time and radius that govern the physical properties of the cosmological Melvin spacetimes.  First is the time dependent factor
\begin{equation}\label{function}
t^{1-a/a_c}
\end{equation}
with $a_c=1/\sqrt{3}$ which appears in both the metric and gauge potential in (\ref{cosmomelvin}).  The impact of this factor depends on whether the dilaton coupling $a$ is less than, greater or equal to the critical value $a_c$.  For $a=a_c$ the function (\ref{function}) equals one and the magnetic field and its impact on the metric in (\ref{cosmomelvin}) are independent of time.  The only time dependence in the metric is the overall $t^{1/2}$ cosmological expansion, while the metric on a constant time slice is identical to the static Melvin metric (\ref{melvin}) with $a=a_c$.  The effects of the cosmic expansion and the magnetic field on the metric are independent in this case.  For $a<a_c$ or $a>a_c$, on the other hand, the function (\ref{function}) is respectively either uniformly increasing or decreasing in time.

The second function we would like to consider is
\begin{equation}
X(r,t)={(1+a^2)\over 4}B^2r^2t^{1-a/a_c}
\end{equation}
which determines the behavior of the function $f=1+X$ in (\ref{cosmomelvin}).  The function $X(r,t)$ combines the time dependence of (\ref{function}) with the characteristic $B^2r^2$ growth of the static Melvin metric function in (\ref{melvin}).  We can think of the function $X(r,t)$ as describing the strength of the backreaction of the magnetic field on the metric.  In the range of coordinates such that $X(r,t)\ll 1$, this backreaction is small and the metric approaches the $w=1$ scalar cosmology (\ref{4milne}).  If we assume that $a<a_c$ and follow the cosmological evolution back to earlier and earlier times, we see that the region with $X(r,t)\ll 1$ will extend out to increasingly large radius, yielding a more and more extended region with metric approaching the pure cosmological one in (\ref{4milne}).  Conversely, for $a>a_c$ this same effect occurs for late times.  We see, therefore, that the cosmological Melvin spacetimes with $a<a_c$ ($a>a_c$) asymptotically approach the pure $t^{1/2}$ cosmology in the far past (future).

\begin{SCfigure}
%\begin{figure}[h]
%\begin{center}
\includegraphics[width=0.60\textwidth]{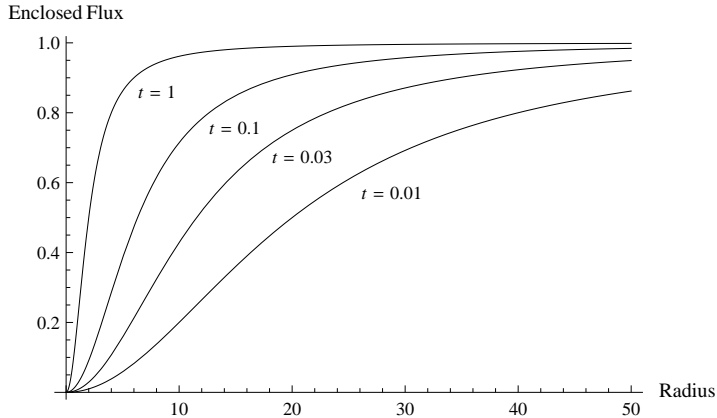}
%\end{center}
\caption{
Magnetic flux $\Phi/4\pi$ enclosed by a circle of comoving radius $r$ for the cosmological Melvin solution with dilaton coupling $a=0$ and $B=1$ for times $t=0.01, 0.03, 0.1$ and $1$.  As time increases, the magnetic flux reaches approaches its asymptotic value more rapidly indicating that the magnetic field is becoming more highly concentrated near the axis. }
\label{fluxfig}
\end{SCfigure}

This limiting behavior of the geometry can also be seen in the behavior of the magnetic flux.
The flux $\Phi$ enclosed by a circle of comoving radius $r$ around the $z$-axis at time 
$t$ is given by
\begin{equation}\label{tflux}
\Phi(r,t) = {4\pi X\over (1+a^2)B(1+X)}%{X\over 1+X}
%{\pi r^2 BH(t)\over 1 +X(r,t)}.
\end{equation}
By taking the limit of  large $r$ at fixed time, one sees that the total magnetic flux remains constant over time and equal to its value (\ref{totalflux}) in the static Melvin spacetimes (\ref{melvin}).
The time derivative of the flux is given by
\begin{equation}
\partial_t\Phi = (1-{a\over a_{c}}) {4\pi X\over (1+a^2)Bt(1+X)^2}
%{\pi r^2 BH(t)\over t(1 +X(r,t))^2}.
\end{equation}
Since the second factor in this expression is positive definite, it follows that for $a<a_{c}$ ($a>a_{c}$) the flux through a circle of fixed comoving radius is always growing (shrinking) in time.
We see that for $a< a_{c}$  the magnetic flux is becomes increasingly concentrated near the symmetry axis as time progresses, approaching its asymptotic value (\ref{totalflux}) at smaller and smaller radii.  Conversely at early times, although the total flux was the same, it was more widely dispersed  in space.  This is consistent with the observation above that the geometry of the spacetimes (\ref{cosmomelvin}) with $a<a_c$ approaches that of the pure cosmological spacetime (\ref{4milne}) out to increasingly large radii at early times.
The evolution of the magnetic flux over time is illustrated for dilaton coupling $a=0$ in figure (\ref{fluxfig}) which plots $\Phi/4\pi$ as a function of radius on a number of constant time slices.  An analogous plot for a value of $a>a_{c}$ would look qualitatively similar, but with the ordering of the time slices reversed, showing that as time increased the asymptotic value of the flux is approached more and more gradually.

%%%%%%%%%%%%%%%%%%%%%%%%%%%%%%%%%%%%%%%%%%%%%%%%%%%%%%%%%%%%%%%%%%%%%%%%%%%%%%%%%%%%%%%%%%%%%%%%%%%%%%%%%%%%%
%
%%%%%%%%%%%%%%%%%%%%%%%%%%%%%%%%%%%%%%%%%%%%%%%%%%%%%%%%%%%%%%%%%%%%%%%%%%%%%%%%%%%%%%%%%%%%%%%%%%%%%%%%%%%%%

Faraday's law tells us that an increasing magnetic flux in the $z$-direction must be accompanied by an electric field in the $-\varphi$ direction.  The electric field $E_\varphi$ seen by an observer at fixed comoving spatial coordinate in the cosmological Melvin spacetimes (\ref{cosmomelvin}) is proportional to $F_{\varphi t}$, which is given by
\begin{equation}
F_{\varphi t} = - (1-{a\over a_{c}}) {2 X\over (1+a^2)Bt(1+X)^2}
%{r^2 BH(t)\over 2t(1 +X(r,t))^2}
\end{equation}
and is simply related to the time varying flux via $F_{\varphi t} =-\partial_t\Phi/2\pi$.  One sees that  as expected, $E_\varphi<0$ for $a<a_{c}$, where we have found that the flux through a circle of constant comoving radius is always increasing in time, and vice-versa for $a>a_{c}$.

It turns out to be interesting to examine the magnetic flux (\ref{tflux}) in the region where $X(r,t)\ll 1$ in more detail.  In this regime we can consider the magnetic field as a perturbation around the $w=+1$ cosmology (\ref{4milne}) with scale factor $s(t)=t^{1/2}$ and write the flux as 
\begin{equation}\label{smallflux}
\Phi(r,t) \simeq \pi r^2 B\,  (s(t))^{2(1-a/a_c)}.
\end{equation}
This displays the increase (decrease) in time we that we have noted above for $a<a_c$ ($a>a_c$).  If we now focus on dilaton coupling $a=0$, in particular, we see that the flux is simply equal to the constant $B$ times the proper area of the disk of comoving radius $r$.  This result disagrees with a standard expectation that a magnetic field should disperse in an expanding universe with a constant flux through a disk of fixed comoving radius (see {\it e.g.} \cite{Durrer:2013pga,Kunze:2013kza}).  

This expectation comes about in the following way.  Because the source-free Maxwell equations are conformally invariant, any solution in flat spacetime continues to be a solution for a flat FRW background, if as in (\ref{4milne}) one works in terms of the conformal time coordinate.  It is natural, therefore, to start with the solution for a constant magnetic field in flat spacetime
\begin{equation}\label{notime}
A= {Br^2\over 2}d\varphi .
\end{equation}
Considering this to be a test magnetic field in {\it e.g.} the FRW background (\ref{4milne}) one finds that the magnetic flux  through a disk of constant comoving $r$ radius is equal to a constant $\pi r^2 B$ and that the electric field vanishes.  

However, the gauge potential for the $a=0$ cosmological Melvin spacetime (\ref{cosmomelvin}) actually has a different behavior in the  $X(r,t)\ll 1$ where it approaches
\begin{equation}\label{withtime}
A\simeq {Br^2 t\over 2}d\varphi . 
\end{equation}
This is another solution to Maxwell's equations in flat spacetime, but obviously differs from (\ref{notime}) through its linear dependence on time.  This produces a magnetic field that increases linearly in time together with a time independent electric field in the $-\varphi$ direction.  This is an intriguing result.  Einstein-Maxwell theory coupled to a massless scalar field is not an exotic physical system and yet we have here
an exact solution for a magnetic field in a cosmological setting that behaves very differently from standard expectations based on test magnetic fields.  We note that the cosmological Melvin solution with $a=a_c$ does have a guage potential of the form (\ref{notime}) in the $X(r,t)\ll 1$ regime.  However, it is unclear from the present work whether we should expect there to be another solution to the $a=0$ field equations that also reduces to (\ref{notime}) in this regime in agreement with standard lore.

The flux of stress-energy through a cylindrical surface provides a further illustration of the dynamical nature of the cosmological Melvin spacetimes (\ref{cosmomelvin}). 
The stress-energy tensor is the sum of the two contributions $T_{ab}=T_{ab}^{(F)}+T_{ab}^{(\phi  )}$, where
\begin{equation}\label{tabmax}
T_{ab}^{(F)} = {e^{-2a\phi}\over 4\pi} \left(  F_{ac} F_b{} ^ c -{1\over 4} g_{ab} F^2 \right), \quad T_{ab}^{(\phi  )} =  {1\over 4\pi}
\left((\nabla _a \phi) \nabla _b \phi-{1\over 2}g_{ab}(\nabla\phi)^2\right)
\end{equation}
Let $t^a $ and $r^a$ be unit timelike and spacelike vectors in the time and radial directions, and let $h_{ab}$ be the induced metric on a cylinder of comoving radius $r$ at time $t$.
The flux of stress energy across the cylindrer is then given by 
\begin{equation}
- \int d\phi dz\sqrt{h}\, T^a {}_b\, t_a r^b = \int d\phi dz \, r\,t\,  T^t{} _r .
\end{equation}
The stress-energy flux of the Maxwell field in the cosmological Melvin solutions (\ref{cosmomelvin}) may be written as
\begin{equation}
T^{(F)t}{}_{r}  =  -(1-{a\over a_{c}} ) {  b X  \over 2\pi rt^2 (1+X)^{2b +2} } 
\end{equation}
where $b=1/(1+a^2)$.
We see that if $a=a_{c}$ the stress-energy flux through the cylinder is zero, in accordance with the other observations we have made in this case. 
If $a< a_{c}$, on the other hand, then there is a flow of electromagnetic energy inwards, consistent with the increasing concentration of magnetic flux
at late times found above. Finally,  if $a>a_{c}$ then the flow of electro-magnetic energy is directed outwards in accordance with the increasing dispersal of magnetic flux in this case.

The dilaton energy flux is more complicated in its behavior and is given by
\begin{equation}
T^{ (\phi )t}{}_{r}  =    - { a b X  \over  4\pi rt^2 f^{2b +2} } \left( 1-b(a+a_c)(a-{1\over a_c})X\right)
\end{equation}
For $0\le a<1/a_c$ the flux of dilaton stress-energy is inward everywhere. For values of the dilaton coupling $a>1/a_c$ the flux is inward in regions where 
$X(r,t)$ is sufficiently small ({\it i.e.}  at small radius and/or late times), while the flux is outward at large $X$ (for large radius and/or early times). Between these two
flows is a surface on which the dilation flux vanishes.  This is where $X(r,t)=(a^2+1)/(a^2-2a_ca-1)$.  
This radius at which the dilaton flux vanishes is found to be increasing in time.

\section{Accelerating cosmological Melvin spacetime}\label{fast}

When $B=0$ the  cosmological Melvin solutions (\ref{cosmomelvin})  reduce to the massless scalar cosmology (\ref{4milne}) with expansion rate  $\tau^{1/3}$ in standard FRW time.  It is natural to ask whether solutions describing magnetic fields in other FRW cosmologies can be found.  If the expansion  is driven by a scalar field, then such more general cosmologies require an exponential potential term in the action for the scalar field \cite{Halliwell:1986ja}.
In this section, we will see that one such example can again be found by starting in $5D$ and Melvinizing.  We will defer a more complete treatment of Melvin fields with exponential scalar potentials to a future publication.

This time, start in $5D$ with a cosmological constant term included in the action
\begin{equation}
S_5=\int d^5x\sqrt{-g^{(5)}}(R-2\Lambda)
\end{equation}
Dimensionally reducing using the Kaluza-Klein identification of the $4D$ fields in (\ref{reduce}) leads to a $4D$ action with an exponential potential term for the dilaton
\begin{equation}
S=\int d^4x\sqrt{-g}\left(R-2(\nabla\phi)^2-e^{-2\sqrt{3}\phi}F^2 - 2e^{2\phi/\sqrt{3}}\Lambda \right).
\end{equation}
Now take the $5D$ deSitter metric in flat FRW coordinates as the seed metric for the construction,
\begin{equation}
ds_5^2 = -d\tau^2 + e^{2H\tau}(dw^2 +\delta_{ij}dx^idx^j)
\end{equation}
with $i,j=1,2,3$ and $H^2=\Lambda/6$.  Plugging this into the Kaluza-Klein form (\ref{reduce}) and transforming to a conformal time coordinate yields the 
$4D$ fields
\begin{equation}
ds_4^2 =(\pm{ 1\over H t})^3\left(-dt^2 +\delta_{ij}dx^idx^j\right ),\qquad e^{-2\phi/\sqrt{3}}= \pm{1\over Ht},
\end{equation}
where the sign is chosen such that $\pm Ht>0$.  Note that the coordinate range $t<0$ yields an expanding cosmology, while $t>0$ is a contracting phase.
In terms of ordinary, non-conformal FRW time this corresponds to an accelerating expansion with scale factor $a(\tau)\sim \tau^3$
and equation of state $p=w\rho$ with $w=-7/9$ (see the appendix).  

We can now add a Melvin magnetic field using the Melvinizing procedure of section (\ref{melvinizing}) which yields the accelerating Melvin cosmology
\begin{eqnarray}\label{accel}
ds^2 &=& (\pm{ 1\over H t})^3\left( f^{1/2}(-dt^2+dz^2+dr^2)+{r^2 d\varphi^2\over f^{1/2}}\right),\qquad A_\varphi = {Br^2\over 2f}\\
e^{-2\phi/\sqrt{3}} &=& \pm {f^{1/2}\over Ht},\qquad f=1+B^2r^2.\nonumber
\end{eqnarray}
This solution is similar in character to the solution originally found in section (\ref{second}) for  dilaton gravity with coupling $a=a_{c}=1/\sqrt{3}$ and vanishing dilaton potential. 
In both cases the magnetic flux through a circle of fixed comoving radius is constant in time and the electric field vanishes.  We expect, however, that cosmological Melvin solutions with more general exponential potentials for the dilaton will have time dependent fluxes and electric fields.

\section{Conclusions}\label{conclude}

In this paper we have found cosmological versions (\ref{cosmomelvin}) of the dilaton Melvin spacetimes (\ref{melvin}) with expansion driven by a massless scalar field.  Key properties of these spacetimes depend on whether the dilaton coupling is less than, greater than, or equal to the value $a_{c}=1/\sqrt{3}$.   With $a<a_{c}$, for example, the magnetic field becomes increasingly concentrated near the symmetry axis as the universe expands.  Tracing the evolution backwards in time, the magnetic flux becomes increasingly dispersed and the spacetime approaches the homogeneous and isotropic cosmology (\ref{4milne}) over an increasingly large region.  We have examined the $a=0$  solution in the limit where the electromagnetic field may be treated as a test field.  Intriguingly, the expression for the gauge potential then reduces to the time dependent configuration (\ref{withtime}), rather than the time independent configuration (\ref{notime}) usually assumed in treatments of magnetic fields in cosmology.  It is natural to ask whether a cosmological solution including gravitational backreaction of the static field (\ref{notime}) exists as well.
Finally, we have also found an example (\ref{accel}) of a Melvin field in an accelerating cosmological background, driven by a scalar field with a specific exponential potential and coupling $a=\sqrt{3}$ to the electromagnetic field strength.

There should be many interesting generalizations of these solutions.  First, we have restricted our attention to four non-compact dimensions, but clearly similar constructions, jointly applying Milnization and Melvinizing, can be carried out in higher dimensions and with other types of gauge fields yielding cosmological versions of more general types of fluxbranes.  One can also look for solutions in $4D$ and higher dimensions with more general power law expansions, corresponding to more general exponential potentials for the dilaton field \cite{Halliwell:1986ja}.  An appropriate system to work with may be the $U(1)^2$ theory used to  construct multi-black hole solutions in cosmological backgrounds with arbitrary power law expansion rates \cite{Gibbons:2009dr,Chimento:2012mg}, which generalize the deSitter multi-black hole solutions of \cite{Kastor:1992nn} and the `stiff matter' solutions of \cite{Maeda:2009zi}.  An initial $U(1)^2$ cosmological Melvin solution, albeit with a vanishing scalar potential, may be obtained by applying the techniques of section (\ref{construct}) starting with $5D$ Einstein-Maxwell theory and including the Kaluza-Klein gauge field as well.  A further goal would be finding cosmological Melvin solutions in more realistic scalar cosmological models of inflation\footnote{See reference \cite{Astorino:2012zm} for a static generalization of the Melvin spacetime with $\Lambda\neq 0$.}. 

One can also hope to study how the properties of black holes in magnetic fields are modified by a scalar cosmological background.   Such an investigation could start with Melvinizing the single-center extremal black hole solutions in \cite{Gibbons:2009dr}.  However, non-extremal black hole solutions to these theories are not yet known.   Finally, one can ask whether there is any generalization of the charged C-metric to a scalar cosmological background which could then be Melvinized to remove conical singularities in analogy with the non-cosmological case \cite{Ernst4}.

\subsection*{Acknowledgements} The authors would like to thank Lorenzo Sorbo for helpful correspondence and Matthew Headrick for use of his \textit{diffgeo} Mathematica package.

\begin{appendix}\label{app}
\appendixpage
\section{Conformal and FRW time coordinates}
In this paper we have worked primarily in terms of the conformal time coordinate for FRW spacetimes.  Although this is common, it is not how one usually compares expansion rates.
In this appendix we provide the general conversion from conformal to standard FRW time, taking the metric between the forms
\begin{eqnarray}\label{times}
ds^2 &=& -(t/t_0)^{2r}\left(-dt^2 + \delta_{ij}dx^idx^j\right)\\
&=& -d\tau^2 +(\tau/\tau_0)^{2p} \delta_{ij}dx^idx^j\nonumber
\end{eqnarray}
One finds that the time coordinates are related according to
\begin{equation}
\left(t/t_0\right)^{r} = \left(\tau/\tau_0\right)^{r\over r+1},\qquad \tau_0=(r+1)t_0
\end{equation}
so that $p=r/(r+1)$ in (\ref{times}).  This gives, for example, $p=1/3$ starting from $r=1/2$ as in the transformation between the two forms of the massless scalar cosmology in (\ref{4milne}) and (\ref{FRWmilne}), or $p=3$ starting from $r=-3/2$ for the acceleration solution in section (\ref{fast}).

\end{appendix}

\end{document}